\documentclass[twocolumn,showpacs,superscriptaddress]{revtex4}
\usepackage{amssymb}
\usepackage{graphicx}
\usepackage{dcolumn}
\usepackage{bm}

\begin{document}

\title{Aharonov-Bohm interference in quantum ring exciton: effects of built-in electric fields}
\author{M. D. Teodoro}
\affiliation{Departamento de F\'{\i}sica, Universidade Federal de S\~{a}o Carlos, 13.565-905, S\~{a}o Carlos, S\~{a}o Paulo, Brazil.}
\author{V. L. Campo Jr.}
\affiliation{Departamento de F\'{\i}sica, Universidade Federal de S\~{a}o Carlos, 13.565-905, S\~{a}o Carlos, S\~{a}o Paulo, Brazil.}
\author{V. Lopez-Richard}
\affiliation{Departamento de F\'{\i}sica, Universidade Federal de S\~{a}o Carlos, 13.565-905, S\~{a}o Carlos, S\~{a}o Paulo, Brazil.}
\author{E. Marega Jr.}
\affiliation{Instituto de F\'{\i}sica de S\~{a}o Carlos, Universidade de S\~{a}o Paulo, 13.566-590, S\~{a}o Carlos, S\~{a}o Paulo, Brazil}
\author{G. E. Marques}
\affiliation{Departamento de F\'{\i}sica, Universidade Federal de S\~{a}o Carlos, 13.565-905, S\~{a}o Carlos, S\~{a}o Paulo, Brazil.}
\author{Y. Galv\~ao Gobato}
\affiliation{Departamento de F\'{\i}sica, Universidade Federal de S\~{a}o Carlos, 13.565-905, S\~{a}o Carlos, S\~{a}o Paulo, Brazil.}
\author{F. Iikawa}
\affiliation{Instituto de F\'{\i}sica ``Gleb Wataghin'', Universidade Estadual de Campinas, 13083-970, Campinas, S\~{a}o Paulo, Brazil}
\author{M. J. S. P. Brasil}
\affiliation{Instituto de F\'{\i}sica ``Gleb Wataghin'', Universidade Estadual de Campinas, 13083-970, Campinas, S\~{a}o Paulo, Brazil}
\author{Z. Y. AbuWaar}
\affiliation{Department of Physics, University of Jordan, Amman 11942, Jordan}
\author{V. G. Dorogan}
\affiliation{Institute of Nanoscale Science and Engineering, University of Arkansas, Fayetteville, Arkansas 72701}
\author{Yu. I. Mazur}
\affiliation{Institute of Nanoscale Science and Engineering, University of Arkansas, Fayetteville, Arkansas 72701}
\author{M. Benamara}
\affiliation{Institute of Nanoscale Science and Engineering, University of Arkansas, Fayetteville, Arkansas 72701}
\author{G. J. Salamo}
\affiliation{Institute of Nanoscale Science and Engineering, University of Arkansas, Fayetteville, Arkansas 72701}

\date{\today}

\begin{abstract}
We report a comprehensive discussion of quantum interference effects due to the finite structure of excitons in quantum rings
and their first experimental corroboration observed in the optical recombinations. Anomalous features that appear in the experiments are
analyzed according to theoretical models that describe the modulation of the interference pattern by temperature and built-in electric fields.
\end{abstract}

\maketitle

The nanoscale ring structures, or quantum rings (QRs), have attracted the interest of the scientific
community due to their unique rotational symmetry and the possibility
to verify quantum mechanical phenomena.\cite{Ring1,Ring2,Ring3} Among these, the study of Aharonov-Bohm (AB)-like effects
has gained a significant impetus,\cite{MG,PA,SE} and these efforts have gone beyond the original discussion
of the AB interpretation on the nature of electromagnetic potentials and their role in quantum mechanics.\cite{AB1}
It is reasonable to say that the study of coherent interference occurring in transport properties of nanoscopic QRs,
as proposed in Ref.~\onlinecite{AB1} encounters, at the moment, serious scale limitations which has encouraged the search for optical implications
associated to AB-effects.

These endeavors applied to nanoscopic QRs do not strictly meet the original
conditions for the AB-configuration since the carriers
are confined within regions with finite values of magnetic field. However,
we still consider an observed effect as of AB-type if it can be explained
assuming that the magnetic field is ideally concentrated in the middle of the
QRs, i.~e., when such effect comes essentially from potential vector-mediated
quantum interference. As also considered in Ref.~\onlinecite{AB2}, in stationary
systems this interference is generally reflected in a boundary condition and
it is not as explicit as in the famous picture of an AB scattering situation.

In this work we consider AB-interference in excitonic states as proposed
theoretically in Refs.~\onlinecite{AC,RA,VIVALDO}. Instead of looking only at 
the oscillatory dependence on magnetic flux of the electron-hole ($e-h$)
recombination energy during photo-luminescence (PL), we also consider the
excitonic oscillator strength whose oscillatory behavior reflects directly the
changes in the exciton wavefunction as the magnetic flux increases. A similar
experimental work was reported in Ref.~\onlinecite{SE} for type-II QRs, however,
here we study type-I systems where both electron and hole move in the ring so that the
correlation between them is crucial to the oscillatory behavior found in the
PL integrated intensity.


\begin{figure}[tbp]
\includegraphics[scale=0.4]{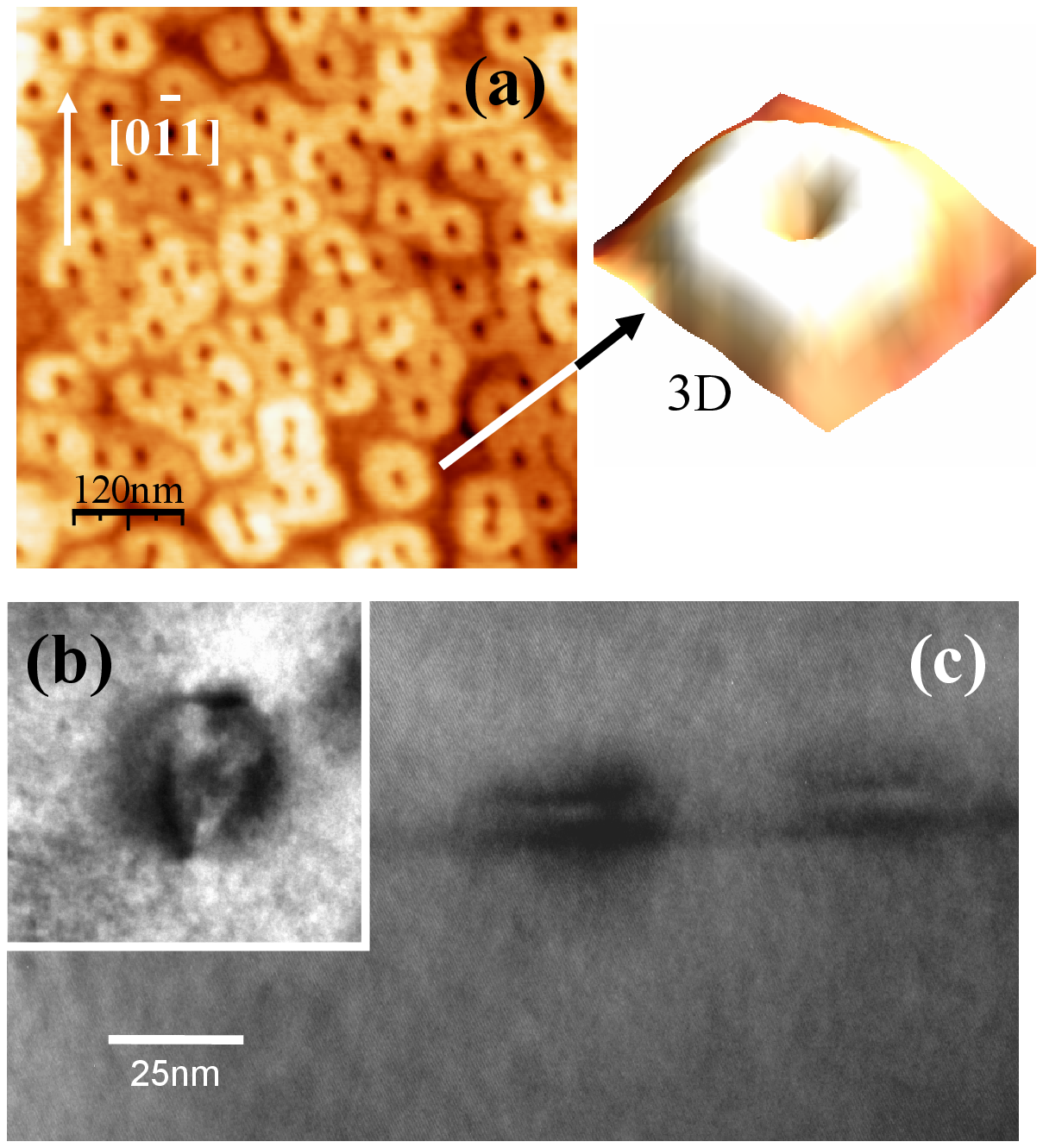}
\caption{(a) AFM image of an uncapped $InAs$ QR sample (the inset shows a 3D profile). Transmission Electron Microscopy images with the same scale: (b) $InAs$ QRs plan-view and (c) cross-section, within the same scale.}
\label{afm}
\end{figure}

The samples studied here were grown using a RIBER 32P solid-source molecular beam epitaxy chamber and the QRs were grown using the following procedure.  A 0.5 $\mu$m $GaAs$ buffer layer was grown on semi-insulating (100) $GaAs$ substrates at 580~$^\circ$C, after oxide desorption. Then, it was followed by 2.2 ML of $InAs$ and the formation of quantum dots (QDs) at 520~$^\circ$C. The dots were obtained using the Stranski-Krastanov growth mode. Cycles of 0.14 ML of $InAs$ plus a 2~s interruption under $As_{2}$ flux were repeated until the total 2.2 ML of $InAs$ was deposited. Next, the QDs were annealed for 30~s to improve their size distribution. The evolution of the dots was detected using in-situ reflection of high-energy electron diffraction. A comprehensive and detailed description of the formation of these sets of self-assembled QRs can be found in Ref.~\onlinecite{EUCLYDES}. Finally, the dots were partially covered  with 4~nm of a $GaAs$ cap layer grown at 520~$^\circ$C to produce the QRs. $InAs$ and $GaAs$ growth rates were set to 0.065 and 1 ML$/$s, respectively.

Atomic Force Microscopy (AFM) was used to investigate the
surface morphology of the grown nanostructures as depicted in \ref{afm}~(a), where an uncapped
sample of $InAs$ QRs is shown with a typical density of about $2.4\times 10^{10}$~cm$^{-2}$. The extracted profile of a single QR is shown in the inset.
To study the optical properties of these nanostructures, the sample was capped with a 50nm $GaAs$ layer .
For the capped sample, morphological and structural analysis were carried out using a FEI Titan
80-300 Transmission Electron Microscope (TEM) equipped with an image
Cs-corrector. TEM samples were prepared using standard mechanical
polishing followed by dimpling and low-energy ion-milling using
Fischione equipment. Figures 1 (a) and (b) show a plan-view and cross-sectional TEM images
that show clear ring-like shapes. Figure 1 (b) displays a plan-view bright-field TEM image of a ring,
taken close to $[001]$ zone axis with ${220}$ reflecting planes. The
density and the diameter of the ring population can be directly
determined from such images. Additional high-resolution imaging, \ref{afm} (c), was performed on cross-section samples that provided
the ring thickness and also confirmed that they perfectly lie on $\{001\}$ planes.
Magneto-PL experiments were performed at 2 K, with magnetic field up to 15 T, using a laser line of $\lambda=$532 nm. The luminescence was detected by a liquid nitrogen-cooled $InGaAs$ diode array.

To formulate a consistent picture of correlation between AB
interference and $e-h$ interaction, we have used a simplified model
according to which the two particle Hamiltonian of the $e-h$ pair
interacting by means of a contact potential in a ring of radius $R$
and zero width threaded by a magnetic flux $\Phi =B\pi R^{2}$ is described
by \cite{VIVALDO}
\begin{equation}
\widehat{H}_{0}=\frac{(\widehat{P}_{e}+\frac{e\Phi}{2\pi Rc})^{2}}{2m_{e}}+\frac{(%
\widehat{P}_{h}-\frac{e\Phi}{2\pi Rc})^{2}}{2m_{h}}-2\pi V\delta (\theta
_{e}-\theta _{h}).  \label{ham1}
\end{equation}%
We used the same gauge as in Ref.~\onlinecite{AB1}, where $\theta _{e(h)}$ is the angular position of electron (hole) and
$\widehat{P}_{e(h)}=\frac{\hbar }{iR}\partial /\partial \theta _{e(h)}$. The
attractive interaction between the electron and the hole has been introduced
by a short range term $2\pi V\delta (\theta _{e}-\theta _{h})$. The
parameter $V$ is chosen so that the exciton ground state obtained in this
model fits the reported value of the exciton binding energy. We have
used the $InAs$ exciton binding energies equal to $4.35$ meV.\cite{LANDOLT}

This Hamiltonian becomes separable if we change to center of mass and
relative position coordinates, $\Lambda =\left( m_{e}\theta _{e}+m_{h}\theta
_{h}\right) /M$, $\theta =\theta _{e}-\theta _{h}$, respectively, where
$M=m_{e}+m_{h}$, and $1/\mu
=1/m_{e}+1/m_{h}$. The exciton eigenfunction will be
\begin{equation}
\Psi _{J}(\Lambda ,\theta )=\frac{e^{iJ\Lambda }}{\sqrt{2\pi }}~e^{-i \frac{%
\Phi }{\Phi _{0}}\theta }\chi(\theta ),  \label{wave}
\end{equation}
where $\Phi _{0}=h~c/e$.
Following Ref.~\onlinecite{AC}, the function $\chi $ satisfies
Bloch's theorem and the twisted boundary condition
\begin{equation}
\chi (\pi )=e^{i2p\pi }\chi (-\pi ),  \label{bc}
\end{equation}%
where $p$ can be restricted to the reduced Brillouin zone, $p\in (-1/2,1/2]$.
Note that $|\chi(\theta)|^{2}$ represents the probability density of finding the electron and hole
angular positions differing by $\theta$. The periodicity of $\Psi (\Lambda ,\theta )$ in $\theta _{e}$ and $\theta _{h}$
determines that the center-of-mass angular momentum, in units of $\hbar $,
assumes integer values $J=0,\pm 1,\pm 2,...$ and that $\gamma J+2(\Phi /\Phi _{0}-p) \in \mathbf{Z}$
(where $\gamma =(m_{h}-m_{e})/M$).\cite{AC} Thus, given an integer value for $J$, the parameter $p$ can be determined uniquely
in the interval $(-1/2,1/2]$. Once we have
determined $p$, we can solve the eigenvalue problem for $\chi $
under the boundary condition (\ref{bc}). The corresponding eigenfunctions can be determined exactly and the exciton energy obtained from a set of 
transcendental equations, as reported in Ref.~\onlinecite{VIVALDO}.

\begin{figure}[tbp]
\includegraphics[scale=0.5]{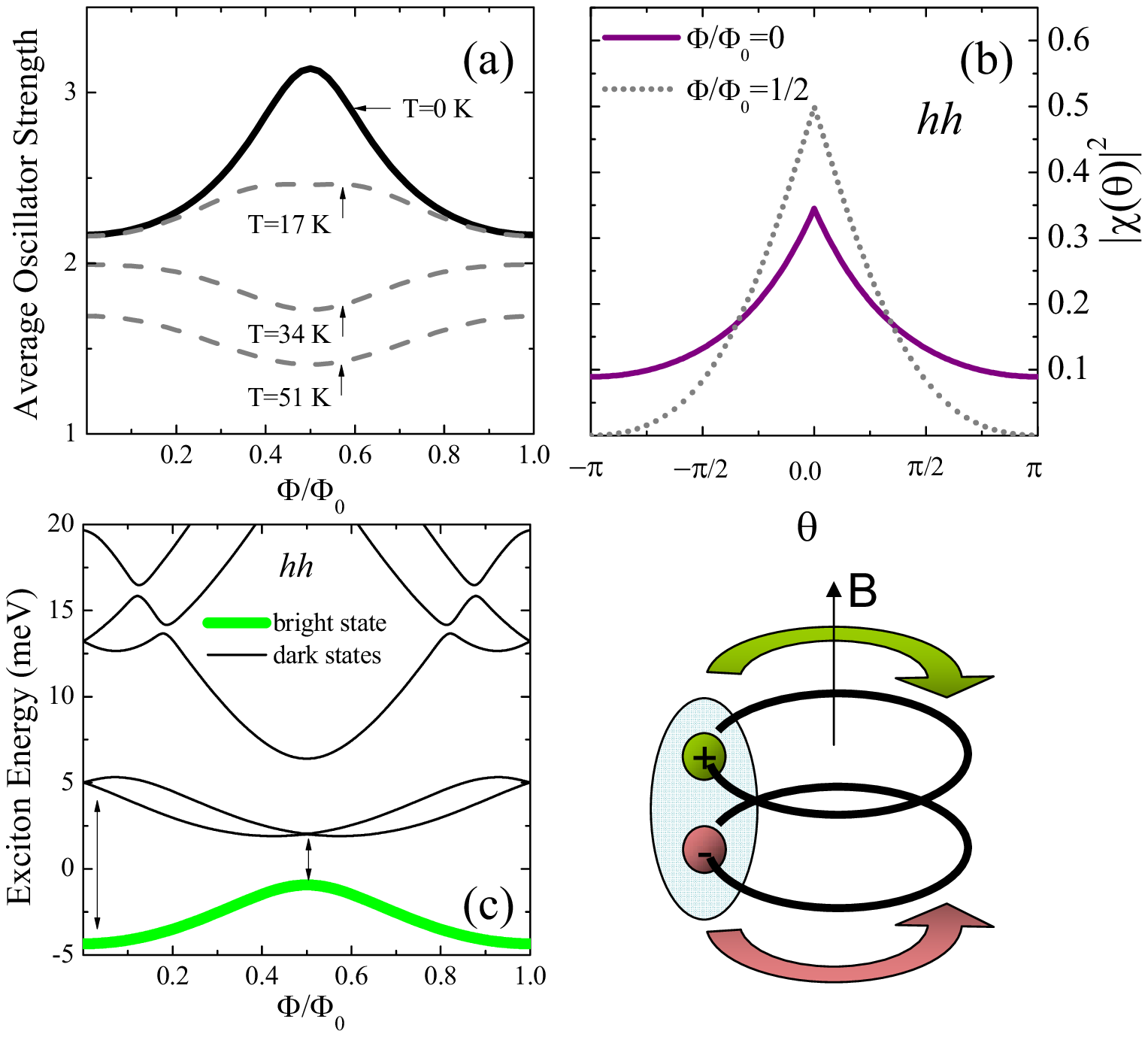}
\caption{(a) Oscillator strength as a function of magnetic flux fraction for a $hh$-exciton at various temperatures. (b) Probability density of finding the electron and $hh$ angular positions differing by $\protect\theta$: the continuous curve corresponds to $\Phi=0$ and
the dotted curve corresponds to $\Phi=\Phi_0/2$. The diagram below represents the $e-h$ interference once
they form a bound pair. (c) Magnetic flux dependence of the first $hh$-exciton states with binding
energy 4.35 meV: the light curve corresponds to the bright exciton while black curves correspond to dark-ones.}
\label{teoria_fig1}
\end{figure}

At this point, it is easy to understand that several physical properties of
this system will be invariant when the magnetic flux through the ring
changes by a multiple of the quantum flux such as the total exciton energy.
Moreover, we can examine the behavior of the oscillator strength that can be unambiguously related to the
relative motion of the electron and the hole given by the function $\chi$ in Eq.~(\ref{wave}).
At T=0 K, the oscillator strength of the ground-state, $J=0$, is given by
\begin{equation}
I_{0}=|\int_{0}^{2\pi }\Psi _{0}(\Lambda ,0)~d\Lambda |^{2} =
\frac{|\chi(0)|^2}{2\pi},  \label{os}
\end{equation}%
therefore is a periodic function on the magnetic flux, as represented in \ref{teoria_fig1} (a).
In the calculations, the $InAs$ band parameters used were: the in-plane masses $m_{h}=(\gamma_{1}+\gamma_{2})^{-1}$ and $m_{e}=0.026$, where $\gamma_{1}=20.4$, and $\gamma_{2}=8.3$.\cite{LANDOLT} Note that we used the heavy-hole ($hh$)-mass in the $(100)$ plane, assuming that the light-hole ($lh$) exciton, with mass $m_{h}=(\gamma_{1}-\gamma_{2})^{-1}$,occupies a higher energy position due to strain effects. Some effects related to the $lh$-excitons will be discussed when necessary.
In order to understand the oscillations, as shown in \ref{teoria_fig1} (a),  in terms of AB-like interference,
we have displayed the function $|\chi|^{2}$ for zero flux and
$\Phi /\Phi_{0}=1/2$, in \ref{teoria_fig1} (b). In both cases, $|\chi|^{2}$ is
$2\pi$-periodic but for $\Phi /\Phi_{0}=1/2$, the exciton wavefunction $\chi$
satisfies anti-periodic boundary condition (Eq.~(\ref{bc})) yielding
the cancelation of the probability density at $\theta=\pm \pi$, which emulates an enhancement of the wavefunction confinement and subsequently its concentration near $\theta=0$. Without $e-h$ interaction the correlation between the electron and hole disappears and the oscillations of the oscillator strength vanish. Thus, the Coulomb binding, although relatively weak due to its short range character, leads to a rather peculiar AB-interference detected optically as oscillations of the PL emission intensity, as suggested in Ref.~\onlinecite{RA}.

\begin{figure}[tbph]
\includegraphics[scale=0.4]{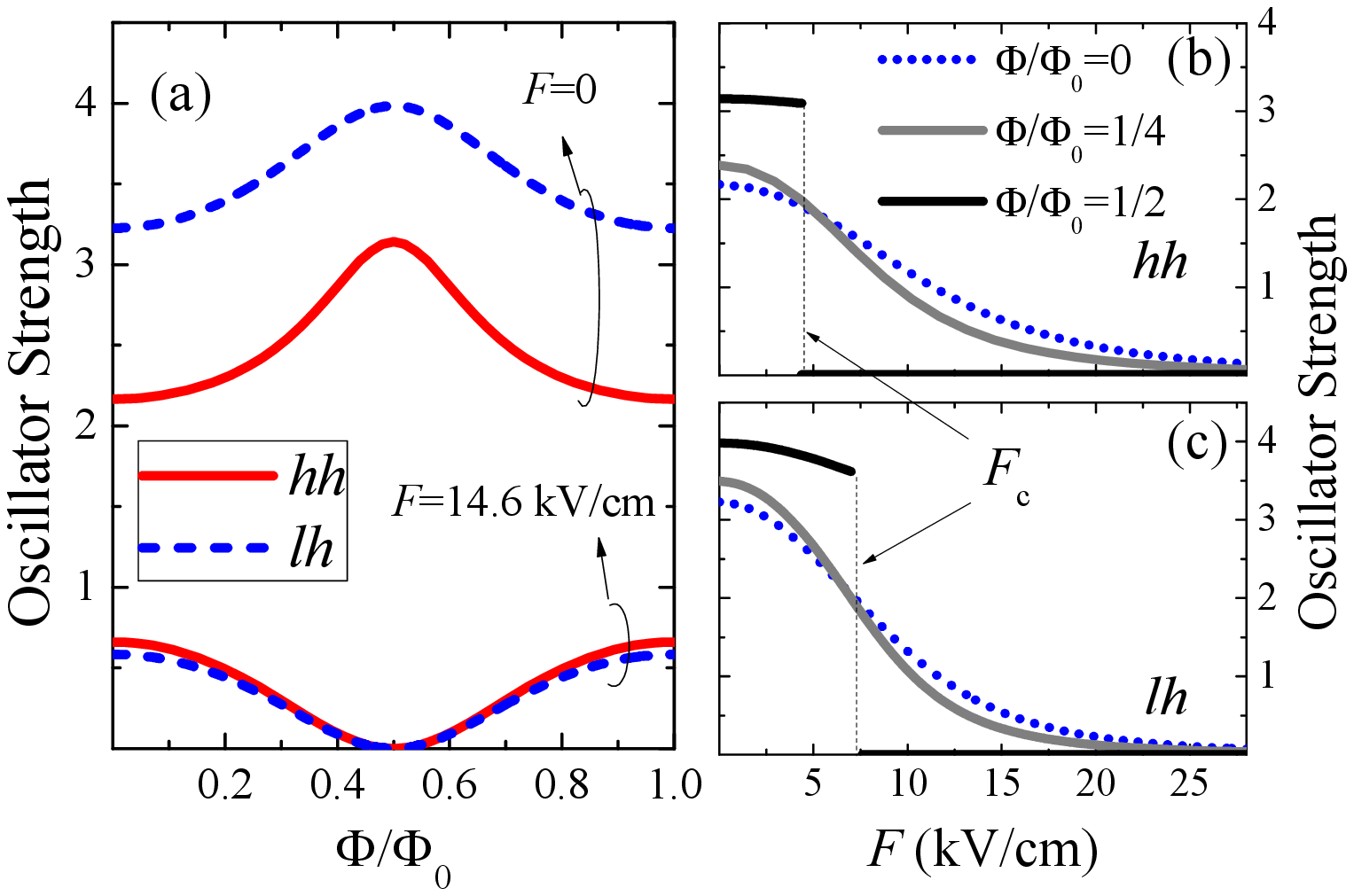}
\caption{(a) Oscillator strength for the exciton in the ground-state as a
function of the magnetic flux at two values of an in-plane electric field ($F=0$ and $F=14.6$ kV/cm):
the solid curve corresponds to the $hh$-exciton, and the dashed curve to the
$lh$-exciton. (b) and (c) Oscillator strength for the $hh$- and $lh$-exciton, respectively, as a
function of the electric field for some values of magnetic flux.}
\label{teoria_fig2}
\end{figure}

We may analyze two elements that invert the sequence of minima and maxima of the oscillator strength: the effect of (i) temperature and (ii) built-in electric fields in the QR plane. In order to take into account the temperature, we generalize Eq.~(\ref{os}) to include all excitonic states,
\begin{equation}
I(T)=\frac{\sum_{J}I_{J}e^{-E_{J}/k_{B}T}}{\sum_{n}e^{-E_{J}/k_{B}T}},
\end{equation}%
where $I_{J}$ is the oscillator strength for the $J-$th state with
wavefunction $\Psi _{J}$, $I_{J}=|\int_{0}^{2\pi }\Psi _{J}(\Lambda ,0)~d\Lambda |^{2}$.
As the temperature rises, the occupation of excited levels of the exciton
becomes more probable.\cite{LIN} In \ref{teoria_fig1} (c) the exciton energy levels are displayed
for $hh$-excitons where the first excited levels correspond
to dark excitons. By increasing the flux, starting from $\Phi /\Phi _{0}=0$
up to $\Phi /\Phi _{0}=1/2$, the dark exciton levels approach the bright-one.
At finite temperature, the net occupation of the ground-state decreases
as these levels come closer and reduces the thermalized
oscillator strength and its amplitude. At relatively high temperatures, this
effect transforms the maximum of the oscillator strength at
$\Phi /\Phi _{0}=1/2$ into a minimum for the $hh$-exciton, as displayed in
\ref{teoria_fig1}(a). 

Another effect that can lead to the inversion of the maximum of the
oscillator strength for the exciton ground-state is the existence of an
in-plane electric field. This internal field can be caused by uniaxial
strains and the subsequent piezoelectricity that can be relatively strong
in $InGaAs$ self assembled 0D structures.\cite{PIEZO1,JA} The in-plane electric field $%
F$\ along $x$-axis can be introduced into the Hamiltonian as
\begin{equation}
\widehat{H}=\widehat{H}_{0}+eFR(\cos (\theta _{e})-\cos (\theta _{h})).
\label{F}
\end{equation}
In order to solve this problem, we adopted a basis set of eigenfunctions
of $\widehat{H}_{0}$ to expand the new solutions. It is easy
to show that the electric field can only couple states with center of mass
angular momenta differing by one. In general terms, we can expect that the oscillator strength will decrease with
increasing electric field, since the electron and the hole are pushed appart.
In \ref{teoria_fig2} (a), the oscillator strength has been displayed as
a function of the magnetic flux for both $hh$- and $lh$-excitons at two values of the in-plane electric field
for T=0 K. It is obvious the inversion of the maximum. In turn, \ref{teoria_fig2} (b) and (c) illustrate the oscillator strength for the
exciton in the ground-state as a function of the electric field. A critical value of the field, $F_{c}$, appears for both $hh$- and $lh$-excitons
where the oscillation pattern becomes inverted. The stronger the electric field, the higher is the projection of
the ground-state over states with odd angular function, $\chi$. This is particularly critical at $\Phi /\Phi _{0}=1/2$ beyond $F_{c}$, when $\chi$ becomes a pure odd function and $\chi (0)=0$. In our QR model, this condition leads to null oscillator strength at $\Phi /\Phi _{0}=1/2$. Certainly, for a QR with finite size, this extreme reduction is not expected.


\begin{figure}[tbp]
\includegraphics[scale=0.6]{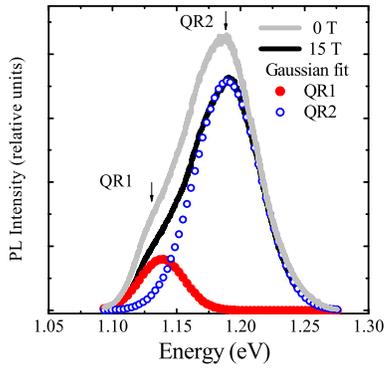}
\caption{Photoluminescence spectra of $InAs$ rings, for $B = 0$ T (gray line) and $B = 15$ T (black line). Both spectra show two radiative channels (filled circles, QR1, and open circles, QR2) identified as the recombination from two sets of QRs of different sizes.}
\label{fig4}
\end{figure}

\begin{figure}[tbp]
\includegraphics[scale=0.3]{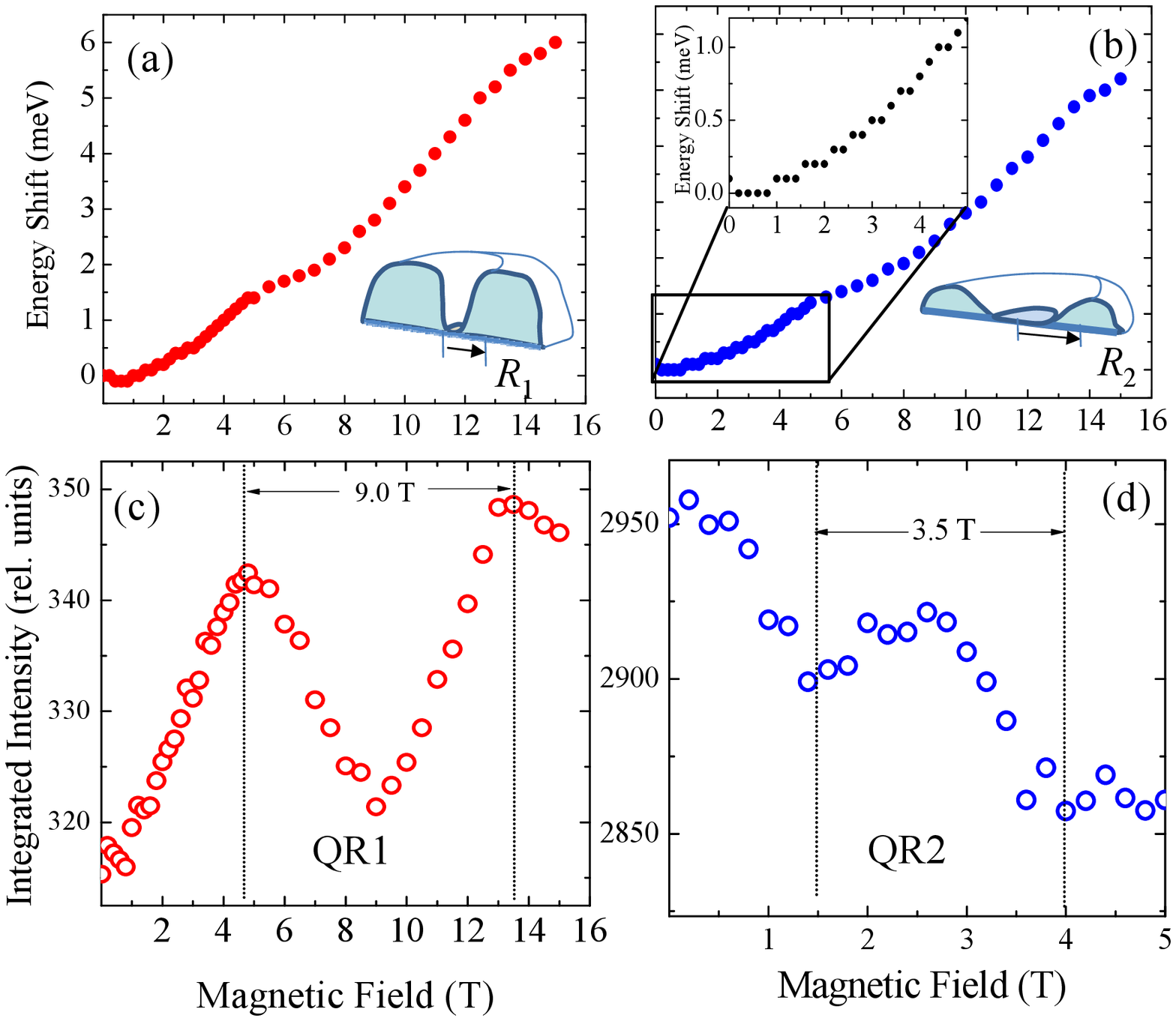}
\caption{
PL peak position energy vs. magnetic field: (a) emission band QR1 (b) QR2. The oscillations of the energy position of the lower energy peak, QR1, can be unambiguously attributed to the quantization of the angular momentum in a QR with finite width.\cite{CH} (c) and (d) show the integrated PL intensity of  QR1 and QR2, vs. magnetic field respectively. The insets in upper figures illustrate diagrams of the rings.}
\label{oscillation_fig}
\end{figure}

PL peak position energy vs. magnetic field: (a) emission band QR1 (b) QR2. (c) and (d) show the integrated PL intensity of  QR1 and QR2, vs. magnetic field respectively. The insets in upper figures illustrate digram of the rings.

We performed experiments in self-assembled $InAs/GaAs$ QRs structures searching for confirmations
of the AB-interference just described. The $InAs/GaAs$ QRs investigated here present a broad optical emission band ($\sim 50$ meV) measured at 2 K, as shown in \ref{fig4}, and a weak shoulder ($\sim 40$ meV) in the low energy range, labeled QR1 and QR2, respectively. We used Gaussian functions to fit two emission bands (circles in \ref{fig4}) attributed to the bi-modal QR distributions dominated, in this case, by high energy QRs. Under an external magnetic field applied along the growth direction a blue shift is observed for both emission bands. PL peak position and its integrated intensity are plotted in \ref{oscillation_fig}. The apparent energy oscillation (less than 0.5 meV) on the diamagnetic shift observed in both bands is attributed to the systematic fluctuation obtained in the fitting process, therefore, it was not considered in this work. However, the oscillation of the PL intensity (\ref{oscillation_fig} (c) and (d)) is a real behavior. We observed opposite direction of the oscillations for two emission bands as the field is increased, minima and maxima occurs in opposite sequence within different field range and period. The PL intensity oscillation of the band QR1 is in good agreement with the modulation of the oscillator strength shown in \ref{teoria_fig1} considering an average QR radius of 11.6 nm. Therefore, it is attributed to the effect of the exciton interference in a ring-like system resulting in an oscillation on the oscillator strength in the presence of an external magnetic field. The opposite behavior observed for QR2, on the other hand, is rather compatible with the model including an electric field shown in \ref{teoria_fig2}, the presence of the electric field inverts the direction of the oscillator strength and results in an average ring radius of 19 nm. The fact that the emission band QR2 has higher transition energy and larger average radius than QR1 suggests that QR2 may present smaller ring width and height than those for QR1.

\begin{figure}[tbp]
\includegraphics[scale=0.9]{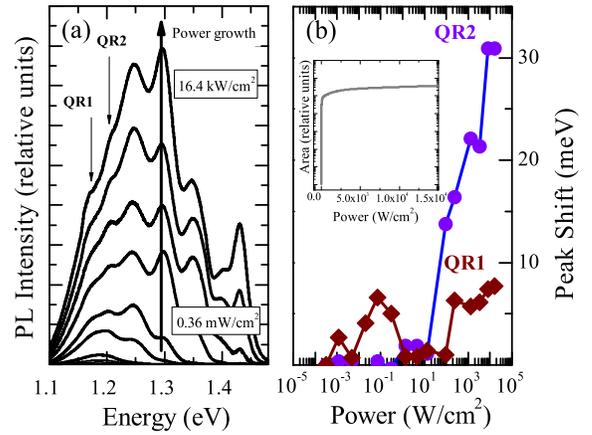}
\caption{(a) Series of the PL emission with the excitation power ranging the interval 0.36 mW/cm$^{2}$-16.4 kW/cm$^{2}$. (b) Position of emission peaks as a function of the excitation power. The inset shows the saturating behavior of the integrated intensity under the QR1 emission band as a function of the excitation power.}
\label{pl_power_fig}
\end{figure}
The effects of built in electric fields in the rings mentioned above has already been described in previous works \cite{JA} when the ring is slightly elongated. During the QR growth anisotropic strain fields are formed in the (001) plane leading to an elongation of the ring in the $[1\overline{1}0]$ direction.\cite{EUCLYDES} This effect, which has been observed in $InAs/GaAs$ QRs, yields a built-in electric field due to the piezoelectric effect.
Additional optical experiments show the evidence of the presence of built-in electric fields in QR2 and negligible field values in QR1, which is consistent with the models discussed above. Figure 6 shows the PL spectra vs. excitation intensity. Increasing the excitation intensity we observed additional emission bands in high energy side region, which are attributed to the state filling of the excited states of the QR2 (dominant band). The main point of this experiment is that, as the carrier population increases in the QRs, the built-in electric field should be screened. In \ref{pl_power_fig} (b) we plotted the energy shift of the QR1 and QR2 versus excitation intensity. We observed a blue shift of the emission band QR2 at the high excitation regime, suggesting the screening of the built-in electric field. The blue shift is not observed in the emission band QR1 where its peak energy remains practically constant confirming thus that the built-in electric field in those QRs is negligible. This result also corroborates the theoretical model used to explain the intensity oscillation of the magneto-optical emissions of QRs. However, the unclear point is why the built-in electric field related to the band QR1 is negligible, since the effect of the elongation should be applied for all rings. A possible reason is that the ring width corresponding to QR1 is larger than that for QR2, as mentioned above and the strain field relief for the former may be stronger than for the latter, thus resulting in a much lower piezoelectric field. Therefore we can attribute the sequence of minima and maxima observed in the intensity oscillations of QR2 to the effect associated to AB-interference combined with the in-plane electric field.


\begin{figure}[tbp]
\includegraphics[scale=0.9]{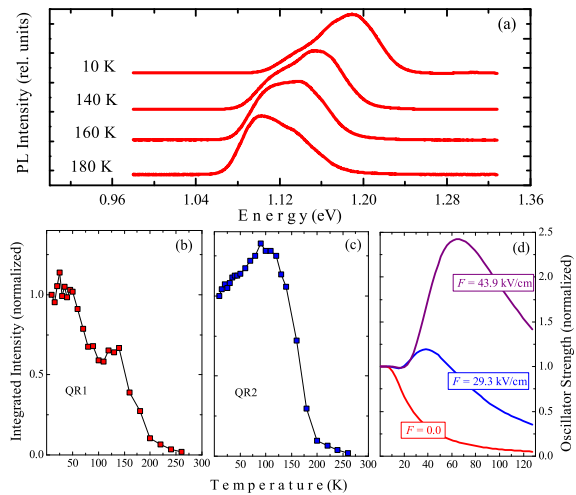}
\caption{(a) PL spectra as a function of temperature. (b) and (c) QR integrated intensity normalized to the value at T=10 K, for the emission bands QR1 and QR2, respectively. (d) calculated oscillator strength (normalized to the value at T=0) for three values of the in-plane electric field $F$. }
\label{pl_temp_fig}
\end{figure}

We also performed the PL experiments vs. temperature in order to investigate the temperature dependence of the excitonic emissions in QR, as discussed in \ref{teoria_fig1}. Figure 7 (a) shows the PL spectra obtained for several temperatures. At low temperature regime the PL spectra are dominated by the emission band QR2, whereas at temperatures above 160~K, it inverts and becomes dominated by QR1. This inversion, at high temperatures, is attributed to the transference of carriers between QRs, via wetting layer, favoring the lower energy states, similar to effects observed in QD systems.

We plotted in Figure 7 (b) and (c) the integrated intensity for each emission band vs. temperature. The decrease of the PL intensity is expected, as shown in Figure 2, due to the occupation of dark excitonic states. However, we observed an increase of the intensity for QR2 at $T<100$ K. This is another signature of the presence of an in-plane electric field in these QRs as corroborated theoretically by the results shown in \ref{pl_temp_fig} (d). The oscillator strength increase occurs by the activation of more efficient channels for optical recombination at excited states, where the $e-h$ angular separation provoked by the electric field is reduced.

The thermal activation energies obtained from Ahrrenius plots could give additional information on the carrier transference. We obtained two activation energies from emission band QR1. One of them at low energy of 13 meV, close to the energy separation between dark and bright (ground) exciton states shown in Figure 2, and the second one, of ~160 meV, related to the PL quenching for T > 150 K, observed in both bands and the total one. The latter is approximately the value estimated for the energy separation between QRs and wetting layer. These experimental results confirm that the inversion in the sequence of minima and maxima for the band QR2, in \ref{oscillation_fig} (d), cannot be attributed to effects induced by temperature, since only for $T>150$ K an effective reduction of the ground-state occupation takes place (far above the temperature value where the oscillations in \ref{oscillation_fig} (d) were detected, T=2 K).



In summary, we have observed and characterized by optical methods an AB-effect
in nanoscopic QRs modulated by built-in piezoelectric fields. The
AB-oscillations, traced by patterns of the PL intensity under increasing
magnetic field, are consistent with the presence of two sets of rings with
different radii in our sample. More interestingly, they show that the
oscillation pattern in the PL integrated intensity for the largest radius
rings has maxima and minima alternated relative to the signal from the
smallest radius rings. Such inversion in sequence of the maxima and minima can
be predicted by a simple theoretical model which takes into account the
correlation between the electron and the hole either as a temperature
effect or as an in-plane electric field effect. We have shown that the source of built-in piezoelectric fields is plausible and experimentally
confirmed and, thus, we have discarded the temperature as another possible cause of the inversion since the experiments are done far below the value at which the presence of dark exciton states becomes relevant. In this way, we confirm that piezoelectric fields may have an important role in strained QR systems inducing a modulation to the AB-oscillations. Also, Coulomb correlation is a crucial
factor for the observations reported in this work; without it,
the correlation between electrons and holes vanishes and these oscillations
disappear.

The authors are grateful to the Brazilian Agencies FAPESP and CNPq for
financial support and the National Science Foundation of the U.S. through Grant \# DMR-0520550.


\begin{thebibliography}{99}

\bibitem{Ring1} J. M. Garc\'{\i}a,  G. Medeiros-Ribeiro,  K. Schmidt, T. Ngo, J. L. Feng, A. Lorke, J. Kotthaus and P. M. Petroff, Appl. Phys. Lett., \textbf{71}, 2014 (1997).

\bibitem{Ring2} A. Lorke, R. J. Luyken, A. O. Govorov, J. P. Kotthaus, J. M. Garc\'{\i}a and P. M. Petroff, Phys. Rev. Lett., \textbf{84}, 2223 (2008).

\bibitem{Ring3} F. M. Alves, C. Trallero-Giner, V. Lopez-Richard and G. E. Marques Phys. Rev. B, \textbf{77}, 035434 (2008).

\bibitem{MG} M. Grochol, F. Grosse and R. Zimmermann, Phys. Rev. B, \textbf{74}, 115416 (2006).

\bibitem{PA} P. A. Orellana and M. Pacheco, Phys. Rev. B, \textbf{71}, 235330 (2005).

\bibitem{SE} I. R. Sellers, V. R. Whiteside, I. L. Kuskovsky, A. O. Govorov, and B. D. McCombe, Phys. Rev. Lett. \textbf{100}, 136405 (2008).

\bibitem{AB1} Y. Aharonov, D. Bohm, Phys. Rev. \textbf{115}, 485 (1959).

\bibitem{AB2} Y. Aharonov, D. Bohm, Phys. Rev. \textbf{123}, 1511 (1961).

\bibitem{AC} A. Chaplik and Pis'ma Zh. \'{E}ksp, Teor. Fiz., \textbf{62},
885 (1995)[JETP Lett. \textbf{62}, 900(1995)]

\bibitem{RA} R. A. R\"{o}mer, M. E. Raikh, Phys. Rev. B \textbf{62}, 7045
(2000), Phys. Stat. Sol. (b), \textbf{221}, 535 (2000).

\bibitem{CH} T. Chakraborty and P. Pietil\"{a}inen, Phys. Rev. B, \textbf{50}, 8460 (1994).

\bibitem{VIVALDO} Andrea M. Fischer, V. L. Campo Jr., M. E. Portnoi, and R.
A. R\"{o}omer, Phys. Rev. Lett. \textbf{102}, 096405 (2009).

\bibitem{EUCLYDES} M. Hanke, Yu. I. Mazur, E. Marega, Jr., Z. Y. AbuWaar, G. J. Salamo, P. Sch\"{a}fer, and M. Schmidbauer
Appl. Phys. Lett. \textbf{91}, 043103 (2007).

\bibitem{LANDOLT} Land\"{o}lt-B\"{o}rnstein Comprehensive Index, edited by O. Madelung
and W. Martienssen (Springer, Berlin, 1996).

\bibitem{LIN} C. H. Lin, H. S. Lin, C. C. Huang, S. K. Su, S. D. Lin, K. W. Sun, C. P. Lee, Y. K. Liu, M. D. Yang, and J. L. Shen, Appl. Phys. Lett. \textbf{94}, 183101 (2009).

\bibitem{PIEZO1} A. Schliwa, M. Winkelnkemper, and D. Bimberg, Phys. Rev. B, \textbf{76}, 205324 (2007).

\bibitem{JA} J. A. Barker, R. J. Warburton and E. P. O'Reilly, Phys. Rev. B,
\textbf{69}, 035327 (2004).



\end{thebibliography}
\end{document}